\documentclass[reprint,superscriptaddress,msmath,amssymb,
amsfonts,amsmath,floatfix
aps,
pra,
]{revtex4-1}
\pdfoutput=1

\usepackage{graphicx}
\usepackage{epstopdf}
\usepackage{amssymb}
\usepackage{amsmath}
\usepackage{xspace}
\usepackage{amsfonts}
\usepackage{mathrsfs}
\usepackage{color}
\usepackage{ulem}
\usepackage[usenames,dvipsnames]{xcolor}
\usepackage[colorlinks=true,citecolor=blue,linkcolor=RubineRed,urlcolor=blue]{hyperref}
\usepackage{phfqit}
\usepackage[utf8]{inputenc}
\usepackage{amssymb}

\usepackage{hyperref}
\colorlet{linkcolor}{blue!50!black}
\hypersetup{bookmarksnumbered=false,bookmarksopen=false,bookmarksopenlevel=1,breaklinks=true,pdfborder={0 0 0},colorlinks=true,anchorcolor=linkcolor,citecolor=linkcolor,filecolor=linkcolor,linkcolor=linkcolor,menucolor=linkcolor,runcolor=linkcolor,urlcolor=linkcolor}
\usepackage[all]{hypcap}

\usepackage[nameinlink,capitalize]{cleveref}

\begin{document}

\title{Steering Interchange of Polariton Branches via Coherent and Incoherent Dynamics}

\author{Diego Tancara}
\affiliation{Centro de Investigaci\'on DAiTA Lab, Facultad de Estudios Interdisciplinarios, Universidad Mayor, Chile}
\author{Ariel Norambuena}
\affiliation{Centro de Investigaci\'on DAiTA Lab, Facultad de Estudios Interdisciplinarios, Universidad Mayor, Chile}
\author{Rub\'en Pe\~{n}a}
\affiliation{Universidad de Santiago de Chile (USACH), Facultad de Ciencia, Departamento de F\'isica, Chile}
\author{Guillermo Romero}
\affiliation{Universidad de Santiago de Chile (USACH), Facultad de Ciencia, Departamento de F\'isica, Chile}
\affiliation{Center for the Development of Nanoscience and Nanotechnology, Estaci\'on Central, 9170124, Santiago, Chile}
\author{Felipe Torres}
\affiliation{Center for the Development of Nanoscience and Nanotechnology, Estaci\'on Central, 9170124, Santiago, Chile}
\affiliation{Departamento de F\'isica, Facultad de Ciencias, Universidad de Chile, Casilla 653, Santiago, 7800024, Chile}
\author{Ra\'ul Coto}
\email{raul.coto@umayor.cl}
\affiliation{Centro de Investigaci\'on DAiTA Lab, Facultad de Estudios Interdisciplinarios, Universidad Mayor, Chile}

\begin{abstract}
Controlling light-matter based quantum systems in the strong coupling regime allows for exploring quantum simulation of many-body physics in nowadays architectures. For instance, the atom-field interaction in a cavity QED network provides control and scalability for quantum information processing. Here, we propose the control of single- and two-body Jaynes-Cummings systems in a non-equilibrium scenario, which allows us to establish conditions for the coherent and incoherent interchange of polariton branches. Our findings provide a systematic approach to manipulate polaritons interchange, that we apply to reveal new insights in the transition between Mott Insulator- and Superfluid-like states. Furthermore, we study the asymmetry in the absorption spectrum by triggering the cavity and atomic losses as a function of the atom-cavity detuning and photon's hopping.
\end{abstract}

\maketitle

\section{Introduction}

Quantum networks are promising platforms for the distribution of quantum information \cite{Kimble,Ritter}, quantum transport \cite{Caruso}, and for simulating complex quantum systems \cite{Georgescu,AngelakisBook,Angelakis,Boulier}. These applications require a high degree of control, which depends on the system at hand. In particular, each node in the network can be considered as a cavity QED containing a single two-level atom, which leads to a light-matter based quantum simulator~\cite{AngelakisBook,Angelakis,Hartmann,Noh_2016,Boulier}, see Fig.~\ref{figure1}(a). The Jaynes-Cumming (JC) model \cite{Jaynes} describes the interaction between the atom and the quantized electromagnetic field, thus introducing hybrid light-matter quantum states termed as Polaritons. The latter corresponds to an atom dressed by the cavity field, and each excitation manifold splits into two different branches, that lead to the lower polariton (LP) and upper polariton (UP) states \cite{Angelakis,Boulier,Grujic_2012,Coto_2015}. These polaritons exhibit different behavior in the dispersive regime of light-matter interaction, where the UP and LP shows atomic and photonic behavior, respectively \cite{Irish}. Steering the interchange between UP and LP opens new avenues to study strongly correlated many body systems, that accounts for well-controlled dynamic of quantum phase transitions or quantum transport. \par

In this work, we propose the interchange of polariton branches (IPB) of a single- and two-sites Jaynes-Cummings systems in a non-equilibrium scenario, which allows us to establish conditions for the coherent IPB. In the former case, a coherent external field acting upon the atomic system drives Rabi oscillations between the two polariton branches. We also study the interchange due to time-dependent detuning. In the realistic situation of an open system we find that the dynamics induced by the coupling of the cavity with its reservoir introduces incoherent interchange between the UP and LP. Moreover, the nature of these two branches leads to an asymmetry in the system's absorption spectrum. In the two-sites JC lattice, we investigate the IPB induced by the hopping dynamics. Furthermore, in this two-sites scenario, we show the role of time-dependent detuning on the dynamics of the well known Mott Insulator- and Superfluid-like states. The interplay between the two branches prevents multiple photons absorption returning to the Mott state even for large detuning; and also allows polariton fluctuations between the lower and upper branches. 

This article is organized as follows. In section \ref{Model}, we briefly describe the JC model. In section \ref{Polariton_Exchange} we analyze four different scenarios for the IPB. We focus on the hopping dynamics, coherent transitions due to atomic driving, both cavity and atomic relaxation, and time-dependent detuning. In section \ref{QPT} we investigate the effect of a time-dependent detuning on the non-equilibrium dynamics of the two-sites JC lattice. In section \ref{Conclusions}, we present the final remarks of this work.

\section{The Model}\label{Model}

\begin{figure*}
	\centering
	\includegraphics[width=15 cm]{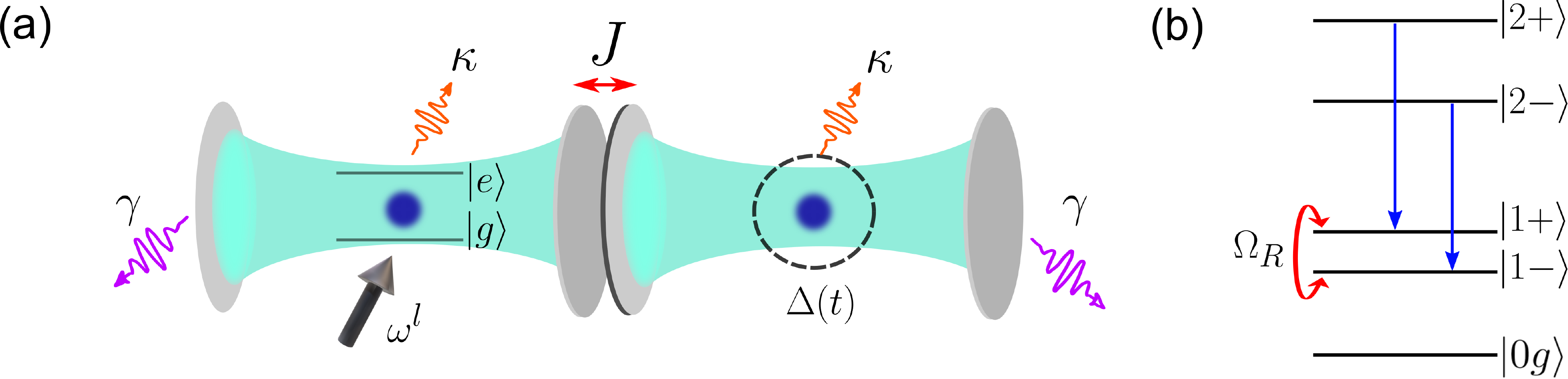}
	\caption{(a) Schematic representation of a cavity QED array with hopping between adjacent sites given by the coupling strength $J_{j}$. (b) Positive and negative branches associated with the Jaynes-Cummings model are given by the states $|n\pm\rangle$ with $n=1,2,...$ being $|0 g\rangle$ the zero energy level. Coherent interchange of polariton branches and branch-preserving decay are highlighted. } 
	\label{figure1}
\end{figure*}

Hybrid light-matter states arise as the eigenstates of the Jaynes-Cumming Hamiltonian \cite{Jaynes}, where different manifolds are separated in terms of the number of photons inside the cavity. For each manifold, two different branches appear, namely the lower and upper branches, as depicted in Fig.~\ref{figure1}(b).

Cavities are connected in a linear array, or even in a more complex network, where photons can hop between nearest-neighbor cavities \cite{Irish,Koch,Coto_2015,Figueroa}. The addition of this hopping leads to the Jaynes-Cummings-Hubbard (JCH) model \cite{Grujic_2012,Hartmann06,Greentree}, which can be described by the Hamiltonian $H=H_{\rm JC}+H_{hp}$, with ($\hbar=1$)
\begin{eqnarray}
H_{\rm JC}  &=&  \sum_{j=1}^{N_{c}} \left ( \omega_j^a \sigma_j^+\sigma_j^- + \omega_j^c a_j^{\dagger}a_j + g_j(a_j^{\dagger}\sigma_j^- + a_j\sigma_j^+)  \right ),\label{Hjc}  \\
H_{hp} &=&  \sum_{j=1}^{N_c-1}(J_j a_j^{\dagger} a_{j+1} +  J_j^\ast a_{j+1}^{\dagger}a_j),\label{Hhp}
\end{eqnarray}
 where $\omega_j^a$, $\omega_j^c$, $g_j$ correspond to the $j$-th atomic frequency, cavity frequency, and atom-field coupling strength, respectively. $a_j^\dagger$ ($a_j$) stands for the creation (annihilation) operator of the $j$-th cavity mode, and $\sigma_j^+=|e\rangle\langle g|$ ($\sigma_j^-=|g\rangle\langle e|$) is the raising (lowering) atomic operator. $N_c$ sets the number of cavities, and $J_j$ corresponds to the hopping strength between neighboring cavities $j$ and $j+1$. The hopping parameter can be tuned in different ways depending on the physical implementation. For instance, it can be achieved through an optical fiber \cite{Eremeev,Coto_2013}, evanescent coupling between the cavities \cite{Greentree,Hartmann,Lepert_2011}, superconducting circuits \cite{Nunnenkamp_2011,Raftery,Fitzpatrick}, and trapped ions \cite{Ivanov,Toyoda}. 

We can write the JCH Hamiltonian ($H$) in the polariton basis by using the following representation of the field and atomic operators \cite{Koch,Coto_2015}
\begin{equation}\label{adagger}
a^{\dagger} =\sum_{n=1}^{\infty} c_{n +}\mathit{L}_{n +}^{\dagger} + \sum_{n =1}^{\infty} c_{n -}\mathit{L}_{n -}^{\dagger} + \sum_{n =2}^{\infty} k_{n \pm}\mathit{L}_{n \pm}^{\dagger} + \sum_{n =2}^{\infty} k_{n \mp}\mathit{L}_{n \mp}^{\dagger},
\end{equation}

\begin{equation}\label{sigma_dagger}
\sigma^+ =\sum_{n=1}^{\infty} c_{n +}^a\mathit{L}_{n +}^{\dagger} + \sum_{n =1}^{\infty} c_{n -}^a\mathit{L}_{n -}^{\dagger} + \sum_{n =2}^{\infty} k_{n \pm}^a\mathit{L}_{n \pm}^{\dagger} + \sum_{n =2}^{\infty} k_{n \mp}^a\mathit{L}_{n \mp}^{\dagger},
\end{equation}
where polariton operators are $\mathit{L}_{n +}^{\dagger}=\ket{n+}\bra{(n-1)+}$, $\mathit{L}_{n -}^{\dagger}=\ket{n-}\bra{(n-1)-}$, $\mathit{L}_{n \pm}^{\dagger}=\ket{n+}\bra{(n-1)-}=(\mathit{L}_{n\mp})^\dagger$. Coefficients $c_{n\pm}$, $c^a_{n\pm}$, $k_{n\pm}$ and $k^a_{n\pm}$ are given in the Appendix \ref{Coefficients}. Following this new representation and considering identical cavities, the JC Hamiltonian can be written in a diagonal form as
\begin{equation}\label{Hjc_pol}
\mathit{H}_{\rm JC} = \sum_{j=1}^{N_{c}} \sum_{n=1}^{N_{f}}(E_{n+}^j\vert n+\rangle_j\langle n+\vert + E_{n-}^j\vert n-\rangle_j\langle n-\vert),
\end{equation}
where $N_f$ is a cut-off excitation number. The eigenstates and corresponding energies are given by
\begin{eqnarray}\label{states}
\vert n -\rangle &=& \cos(\theta_n)\vert n,g\rangle -\sin(\theta_n)\vert n-1,e\rangle ,\\
\vert n +\rangle &=& \sin(\theta_n)\vert n,g\rangle + \cos(\theta_n)\vert n-1,e\rangle ,\\
E_{n\pm} &=& \omega^c n +\frac{\Delta}{2}\pm \frac{\sqrt{\Delta^2 +4g^2 n}}{2}.
\end{eqnarray} 

Here, $\theta_{n}=\frac{1}{2}\arctan(\frac{g\sqrt{n}}{\Delta/2})$, $\Delta=\omega^a-\omega^c$, and $n$ corresponds to the number of photons inside each cavity. One can see that for a fixed number of photons, each subspace splits into a lower (LP) and upper (UP) polaritons, separated by $R_{n}= E_{n+}-E_{n-}= \sqrt{\Delta^2+4g^2n}$. The UP is usually suppressed by preparing a LP initial state and following only resonant transitions \cite{Birnbaum,Angelakis}. Nevertheless, we show here that the UP may appear due to different dynamics.
 
In what follows, we detail the conditions and parameter regimes where one branch can be isolated from the other (no polariton interchange), and the opposite case where Rabi oscillations are observed, i.e., $\vert n-\rangle \leftrightarrow \vert n+\rangle$. For this goal, we consider four different resources, namely: the hopping dynamics, external driving, relaxation due to the interaction with a Markovian environment, and a time-dependent detuning. These are commonly available resources that will allow us to get further control on the system, harnessing light-matter interaction.

\section{Polaritons Interchange}\label{Polariton_Exchange}

\subsection{Fast oscillations in hopping dynamics}\label{section_SRWA}

In this subsection we focus on the hopping Hamiltonian that is responsible for the connection between sites. The interaction between adjacent cavities can be written in the polariton basis as follows,
\begin{eqnarray}\label{Hhp_pol}
\mathit{H}_{hp} &=& \sum_{j=1}^{N_c -1} J_j[(P_{+j}^\dagger + P_{-j}^\dagger + P_{\pm j}^\dagger + P_{\mp j}^\dagger ) \nonumber \\&\times & (P_{+(j+1)} + P_{-(j+1)} + P_{\pm (j+1)} + P_{\mp (j+1)})+\rm{h.c.}], \nonumber\\ 
 \end{eqnarray}
where, for simplicity, we set $J_j$ to be real and we rename each term in Eq. (\ref{adagger}) as $\mathit{P}_+^\dagger$, $\mathit{P}_-^\dagger$, $\mathit{P}_{\pm}^\dagger$ and $\mathit{P}_{\mp}^\dagger$, respectively. The Hamiltonian in Eq.~(\ref{Hhp_pol}) can be simplified by performing the Rotating Wave Approximation (RWA) that neglects the contribution of interchanging products like $P_{+j}^\dagger P_{-(j+1)}$ and $P_{+j}^{\dagger}P_{\pm(j+1)}$ \cite{Angelakis}, provided that $g>4J$ \cite{Koch}. For illustration, we formally derive the oscillating terms of the hopping Hamiltonian in the interaction picture, 
\begin{eqnarray}\label{pmaspmen}
\tilde{P}_{+j}^{\dagger}\tilde{P}_{-(j+1)}&=&\sum_{n=1}^{N_{f}} c_{n+}\mathit{L}_{n+}^{j\dagger} e^{it(R_{n}-R_{n-1})}\nonumber\\&\times&\sum_{n^\prime =1}^{N_{f}} c_{n^\prime -}\mathit{L}_{n^\prime -}^{(j+1)} e^{it(R_{n^\prime}-R_{n^\prime -1})},
\end{eqnarray}
%\begin{eqnarray}\label{pmas1}
%\tilde{P}_{+j}^{\dagger}\tilde{P}_{+(j+1)}&=&\sum_{n=1}^{N_{f}} c_{n+}\mathit{L}_{n+}^{j\dagger} e^{it(R_{n+}-R_{(n-1)+})} \nonumber\\&\times&\sum_{n^\prime =1}^{N_{f}} c_{n^\prime +}\mathit{L}_{n^\prime +}^{(j+1)} e^{-it(R_{n^\prime +}-R_{(n^\prime -1)+})},
%\end{eqnarray}
where $\tilde{P}_{+j}^{\dagger} = U P_{+j}^{\dagger}U^{\dagger}$ with $U=\mbox{exp}(it\mathit{H}_{\rm JC})$. Note that $R_{n}=\sqrt{\Delta^2+4g^2n}$ is only defined for $n\geq 1$, otherwise is zero. For $n=1$ (one photon per cavity), the exponent oscillates with frequency $2R_1$. Hence, in the parameter region where $g>4J$ these oscillating terms can be eliminated by the RWA. For $n>1$, we numerically observe that for $g=10J$, the RWA remains as a good approximation. For example, we set $\Delta=0$, $ \omega^c=10^{4}g$ for a two-sites lattice and observe that for the initial state $\ket{1-,0}$, the probability of finding the UP ( $\ket{0,1+}$) due to the hopping interaction only reaches $p_{0,1+}=0.02$. Now, we extend the calculation for the $n=2$ manifold, starting from the initial state $|2-,0\rangle$, and observe that the probability of finding a UP state like $\ket{1-,1+}$ increases up to $p_{1-,1+}=0.08$, but it is still small. 

It is worth noticing that products $\tilde{P}_{+j}^\dagger \tilde{P}_{+(j+1)}$ and $\tilde{P}_{-j}^\dagger \tilde{P}_{-(j+1)}$ that do not interchange polaritons, cannot be eliminated, as detailed in the Appendix \ref{Appendix1}. Then, these two operators will be the only terms in the hopping that matter during the time evolution. The latter means that from an initial LP state and under a pure Jaynes-Cumming-Hubbard evolution,  UP states never show up. Hence, polaritons interchange can be tuned in the hopping dynamics by appropriately choosing the rate $J/g$. In the next subsection we study additional control via external driving.%Thereafter, we will refer to the Full model as the one described by Eq.~\eqref{Hhp_pol}, and the RWA model after performing the RWA for $g\geq 10J$.

\subsection{External driving}

In this subsection we focus on a single JC system, assuming that each site can be individually addressed.  Since atomic and cavity excitations can be manipulated by optical/microwave external fields, polaritons interchange could be assisted in the same way. For instance, suppose the cavity is driven by a continuous wave with frequency $\omega^p$ and coupling strength $\alpha$, while the atom is driven with frequency $\omega^l$ and Rabi coupling $\Omega$. In this case, the Hamiltonian for a single cavity reads $H=H_{\rm JC} + H_{\rm I}$, with $H_{\rm JC}$ given Eq.~(\ref{Hjc}) for $N_c=1$, and the interaction Hamiltonian reads
 \begin{equation}
 H_{\rm I} =  i\Omega(\sigma^+ e^{-i\omega^l t} - \sigma^- e^{i\omega^l t}) + i\alpha(a^\dagger e^{-i\omega^p t} -a e^{i\omega^p t}).
 \end{equation}
 
 In a multi-rotating frame with the atom and cavity frequencies, we obtain
\begin{eqnarray}\label{hamiltonian_double_rotating}
 \tilde{H} &=& \Delta_a\sigma^+\sigma^- + \Delta_c a^\dagger a + g(a^\dagger\sigma^- e^{i\Delta_1 t} \nonumber\\ &+& \sigma^+ a e^{-i\Delta_1 t}) + i\Omega(\sigma^+ - \sigma^-) + i\alpha(a^\dagger  -a ),
 \end{eqnarray}
where $\Delta_a=\omega^a-\omega^l$, $\Delta_c=\omega^c-\omega^p$ and $\Delta_1=\omega^p-\omega^l$. For convenience, we set $\Delta_1=0$ and write Eq.~(\ref{hamiltonian_double_rotating}) in the polariton basis, 
\begin{eqnarray}\label{hamilt_polar_driving}
 \tilde{H} &=& \sum_{n=1}^{N_{f}}(E_{n+}^0\vert n+\rangle\langle n+\vert + E_{n-}^0\vert n-\rangle\langle n-\vert \nonumber \\ &+& \beta_{n+}(\mathit{L}_{n+}^{\dagger}-\mathit{L}_{n+}) + \beta_{n-}(\mathit{L}_{n-}^{\dagger}-\mathit{L}_{n-}))\nonumber \\ &+& \sum_{n=2}^{N_{f}} (\xi_{n\pm}(\mathit{L}_{n\pm}^{\dagger}-\mathit{L}_{n\mp}) + \xi_{n\mp}(\mathit{L}_{n\mp}^{\dagger}-\mathit{L}_{n\pm})  ),
 \end{eqnarray} 
where the coefficients are: $\beta_{n+}=(i\Omega c_{n+}^a + i\alpha c_{n+})$, $\beta_{n-}=(i\Omega c_{n-}^a + i\alpha c_{n-}) $, $\xi_{n\pm}=(i\Omega k_{n\pm}^a + i\alpha k_{n\pm}) $ and $\xi_{n\mp}=(i\Omega k_{n\mp}^a + i\alpha k_{n\mp})$. 

We now seek for a parameter regime where external driving fields allow us to control interchange of polariton branches (IPB). For convenience, we separately research the weak and strong driving regimes. For weak driving ($\alpha$,$\Omega \ll g$), we treat terms proportional to $\beta$ and $\xi$ as a perturbation. The unperturbed eigenenergies are ($\Delta_1=0$)
\begin{equation}
 E_{n\pm}^{(0)} = \Delta_c n +\frac{\Delta}{2}\pm \frac{\sqrt{\Delta^2 +4g^2 n}}{2}.
 \end{equation}
 
  Perturbative contributions to the eigenenergies and eigenstates are considered up to second order, see Appendix \ref{Appendix3}. Without loss of generality, let us focus on the contributions to $\ket{1-}$ state that allow interbrach transitions to $\ket{1+}$ state,
\begin{equation}
 \ket{\widetilde{1-}} \approx (-\frac{\beta_{1-}\beta_{1+}}{E_{1-}^0} - \frac{\beta_{2-}\xi_{2\mp}}{E_{1-}^0- E_{2-}^0} - \frac{\xi_{2\pm}\beta_{2+}}{E_{1-}^0 - E_{2+}^0})\frac{\ket{1+}}{E_{1-}^0 - E_{1+}^0}. 
 \end{equation}
 
One can see that the transition $\ket{1-} \rightarrow \ket{1+}$ occurs as a second-order process. We found that in the weak driving setting, $\Omega=\alpha=0.1\, g$, and also for $\Delta_c=0$, the only relevant transitions happen inside the initial state branch, e.g. $\ket{1-} \rightarrow \ket{2-}$. Thus, in this regime polariton interchange is not observed, and this enables the isolation of a single branch.  

Beyond the weak driving regime, in an intermediate regime where perturbation theory is no longer valid, $\Omega,\alpha\approx g$, polaritons interchange occurs.
 
\begin{figure}[t]
 	\centering
 	\includegraphics[width=0.45 \textwidth]{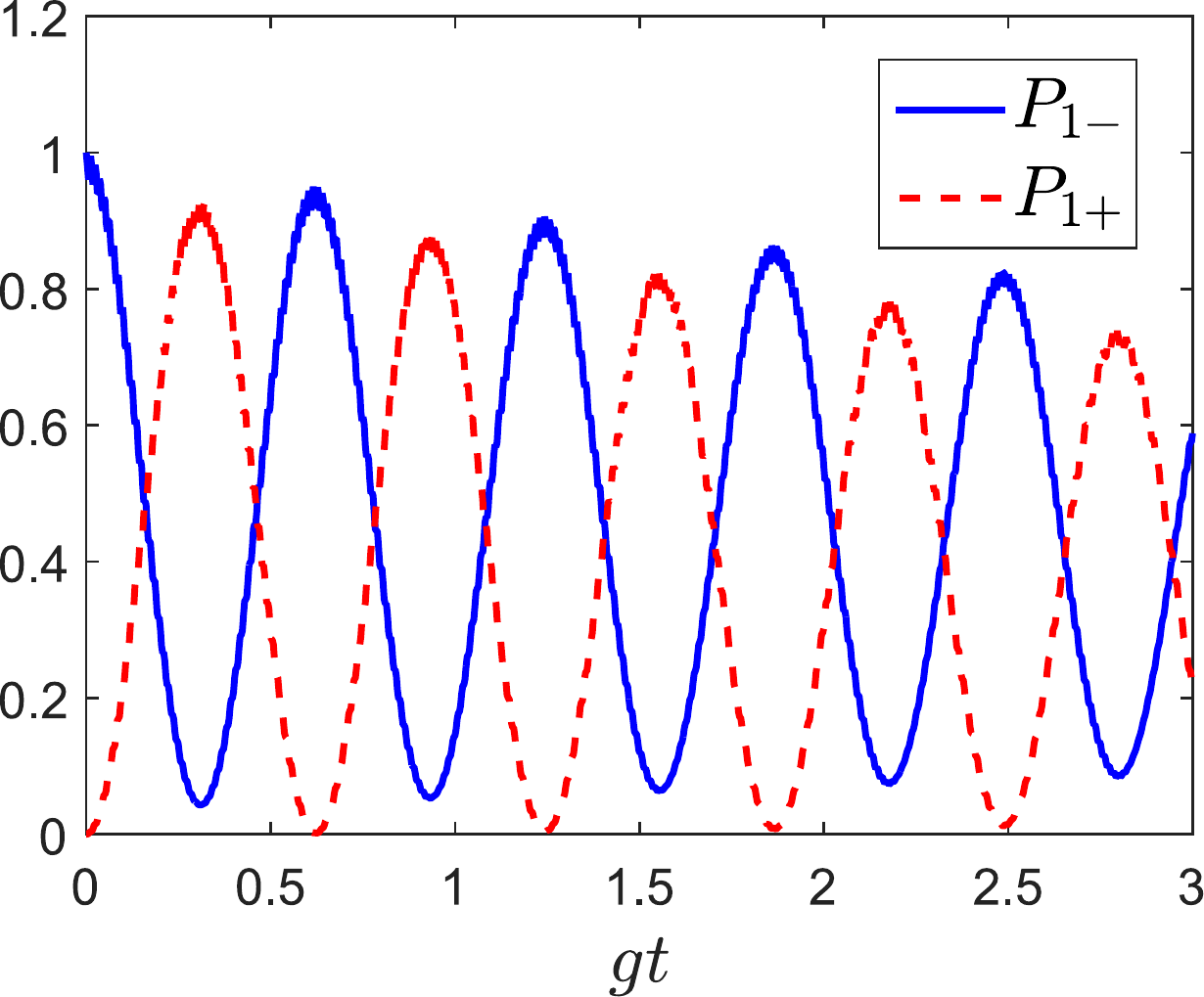}
 	\caption{Second order transitions between polaritons for the $n=1$ manifold are observed in the strong driving ($\Omega=50g$) and large detuning ($\Delta_a=500g$) regime. The period $T=0.627g^{-1}$ is well reproduce with our analytical result for $\Omega_R$ ($T=0.616g^{-1}$). Cavity longitudinal relaxation ($\gamma=0.1g$) only decreases the oscillation amplitude due to the decay to the ground state. Other parameters are $\alpha=0$, $\omega^c=10^4 g$, $\Delta=0$, $\kappa=0$.} 
 	\label{figure3}
 \end{figure}

The most interesting case arises in the strong driving regime. For simplicity, we only consider the atomic driving ($\alpha =0$). For large detuning ($\Delta_a=500g$), this driving induces a second-order process that originates Rabi oscillations between polaritons $\ket{1-}\leftrightarrow \ket{1+}$. We obtain a Rabi frequency, $\Omega_R\approx 2\sqrt{g^2+(\Omega^2/\Delta_c)^2}$. The oscillation period $T = 2\pi/\Omega_R$, is consistent with the one obtained for the dynamics induced by the Hamiltonian in Eq.~\eqref{hamiltonian_double_rotating} for $g=1$, see Fig.~\ref{figure3}. 
  
 Then, we are able to coherently control polaritons interchange, which enables the implementation of quantum gates inside the $n=1$ manifold. We remark that the driven Jaynes-Cumming model has been previously studied in different contexts, such as nonlinear oscillator \cite{Chough} and dissipative phase transition \cite{Carmichael}. In the next subsection we explore in more detail the effects of an open dynamics.  
   
\subsection{Relaxation in Markovian environment}

In a realistic scenario, the system is subjected to relaxation processes due to the interaction with the surrounding environment. This dynamics is commonly modeled by a master equation \cite{Breuerbook}. In this subsection we focus on both cavity and atomic losses, and seek for its representation in polariton operators. Furthermore, we analyze the absorption spectrum for the case of single and double cavity systems. For a single cavity QED, photons decay through imperfect mirrors with a rate $\gamma$ and the atomic excited state experiences spontaneous emission at rate $\kappa$. These combined processes are well described with the Lindblad master equation,
\begin{eqnarray}\label{ME}
\dot{\rho}&=&-i[H_{\rm JC},\rho]+\frac{\gamma}{2}(2a\rho a^\dagger -\lbrace a^\dagger a,\rho \rbrace)\nonumber \\
&& +\frac{\kappa}{2}(2\sigma^{-}\rho \sigma^+ -\lbrace \sigma^+  \sigma^-,\rho \rbrace).
\end{eqnarray}

The above equation hides an interesting detuning-dependent asymmetry that originates from polariton states. Spectral asymmetries have been observed in molecular exciton-polaritons fluorescence for transverse relaxation processes~\cite{Neuman} and in plasmon-polariton systems~\cite{Kongsuwan2019}. For further illustration, we calculate the absorption spectrum of the system,
\begin{equation}\label{AbsSpectrum}
S(\omega) = 2\mbox{Re} \left\{ \int_{0}^{\infty}\langle \langle a(\tau)a^{\dagger}(0)\rangle \rangle_{\rm ss} e^{i\omega \tau}\; d\tau \right\},
\end{equation}

where $\langle \langle a(\tau)a^{\dagger}(0)\rangle \rangle_{\rm ss}$ is the two-point correlation function evaluated at the steady state. In the three-level manifold composed by states $|1\rangle = |1+\rangle$, $|2\rangle = |1-\rangle$, and $|3\rangle = |0g\rangle$, we find an analytical expression for the absorption spectrum ($\hbar = 1$),
\begin{align}
S(\omega) &= 2\sin^2(\theta_1) {\gamma_{+} \over \left(\omega - E_{1+}\right)^2 + \gamma_{+}^2} \nonumber \\ &+ 2\cos^2(\theta_1){\gamma_{-} \over \left(\omega - E_{1-}\right)^2 + \gamma_{-}^2}.\label{AbsSpectrumAn}
\end{align}

We observe that resonances occur at polaritonic energies $E_{1\pm}$, the full-width-at-half-maximum is given by the rates $\gamma_{+} = (1/2)(\sin^2(\theta_1)\gamma +\cos^2(\theta_1)\kappa)$ and $\gamma_{-} = (1/2)(\cos^2(\theta_1)\gamma +\sin^2(\theta_1)\kappa)$ and the amplitudes are modulated by the probability factors $\sin^2(\theta_1)$ and $\cos^2(\theta_1)$. The above expression has been extensively used in atomic, molecular and solid-state systems~\cite{ArielNJP2020}. For more details about its derivation see Appendix~\ref{AbsorptionSpectrum}. In Fig.~\ref{figure2}(a) we show the absorption spectrum for the resonant transitions $|0g\rangle \rightarrow |1\pm\rangle$ calculated for $\Delta = 0$ and $\Delta = g$, using $\gamma = \kappa$. When $\Delta = 0$, the absorption spectrum is symmetric around the cavity frequency $\omega^c$ and the resonant frequencies are $\omega_A = \omega^c-g$ and $\omega_B = \omega^c+g$ for $n=1$. Conversely, for $\Delta = g$ the symmetry is broken and we observe two peaks at frequencies $\omega_A = \omega^c+(1-\sqrt{5})g/2$ and $\omega_B = \omega^c+(1+\sqrt{5})g/2$. From our analytical result given in Eq.~\eqref{AbsSpectrumAn} we note that intensities of the peaks A and B in Fig.~\ref{figure2}(a) are given by $2\cos^2(\theta_1)/\gamma_{-}$ and $2\sin^2(\theta_1)/\gamma_{+}$, respectively. Moreover, since $\gamma = \kappa$, we have $\gamma_{+} = \gamma_{-}$ and the asymmetry originates from the detuning through the polaritonic angles $\theta_1 = (1/2)\mbox{arctan}(2g/\Delta)$.\par
  \begin{figure*}[ht]
	\centering
	\includegraphics[scale=0.42]{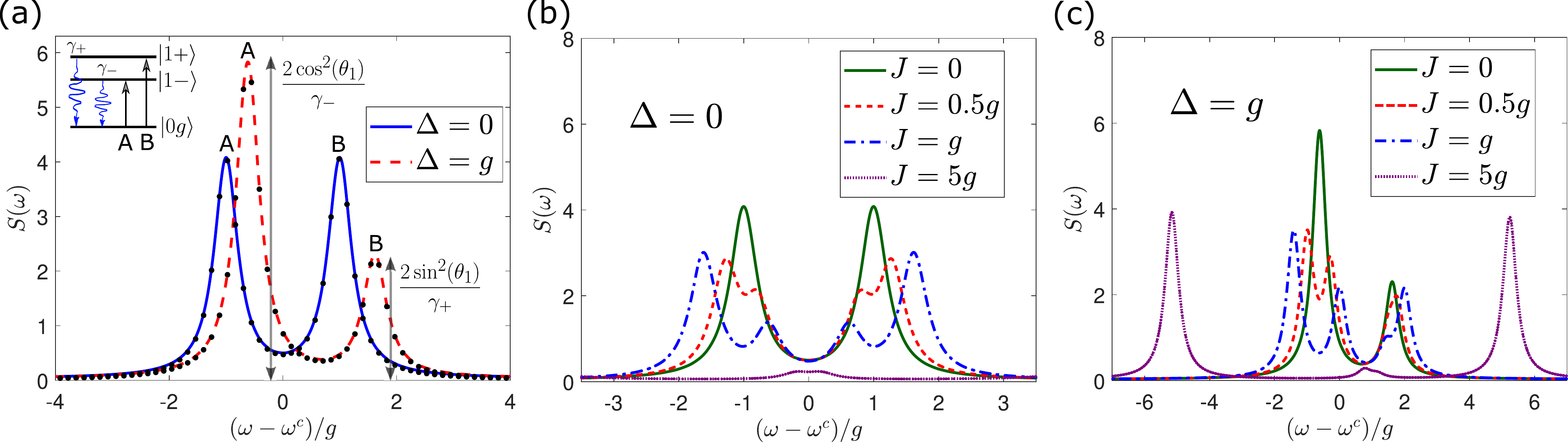}
	\caption{(a) Numerical and analytical absorption spectrum $S(\omega)$ for the resonant transitions A ($|0g\rangle \rightarrow |1-\rangle$) and B ($|0g\rangle \rightarrow |1+\rangle$) with $\Delta = 0$ (solid) and $\Delta = g$ (dashed). Analytical predictions (black circles) from Eq.~\eqref{AbsSpectrumAn} agree with numerical results. For the calculation of the spectrum we use $\omega^c=10^2g$, $\gamma=\kappa = g/2$ and $g = 1$. (b)-(c) Numerical absorption spectrum for two interacting cavities controlled by the hopping strength ($J$) for different detunings. For the simulation of two interacting cavities we use $\gamma_{1,2} = \kappa_{1,2} = g/2$.}
	\label{figure2}
\end{figure*}

For two interacting cavities the master equation reads, 
\begin{align}
\dot{\rho} &= -i[H_{\rm JC} + H_{hp},\rho] + \sum_{j=1}^{2}\gamma_j\left[a_{j} \rho a_j^{\dagger} - {1 \over 2}\left\{a_j^{\dagger} a_j ,\rho \right\} \right] \nonumber\\
&+ \sum_{j=1}^{2}\kappa_j\left[\sigma_{j}^{-} \rho \sigma_j^{+} - {1 \over 2}\left\{\sigma_j^{+} \sigma_j^{-},\rho \right\} \right],\label{me}
\end{align}
 
where $H_{\rm JC}$ and $H_{hp}$ are the Jaynes-Cummings and hopping Hamiltonian for $N_c = 2$. In order to quantify the absorption spectrum, we focus on the first cavity and used Eq.~\eqref{AbsSpectrum}. In Fig.~\ref{figure2}(b)-(c) we plot the numerical absorption spectrum for $\Delta = 0$ and $\Delta = g$ by considering different values for the hopping. For $J = 0$, we recover our previous result for a single cavity. As a consequence of the hopping dynamics ($J \neq 0$), each peak  splits out into two peaks due to the interaction with the second cavity. When $\Delta =0 $, we observe a symmetrical spectrum for each value of $J$. However, for $\Delta = g$, we observe an asymmetrical spectrum in the weak-to-medium coupling regime between the cavities. Furthermore, in the strong coupling regime ($J \gg g$), the spectrum is dominated by the hopping dynamics and its symmetry is restored.
 
In some situations the atom decay can be neglected, for instance using long-lived Rydberg atoms. Even in this case polaritons exhibit detuning-dependent asymmetry, which can be easily found in the annihilation operator $a=\mathit{P}_{-}+\mathit{P}_{+} + \mathit{P}_{\pm} + \mathit{P}_{\mp}$. We remark that coefficients $k_{2\pm}$ and $k_{2\mp}$, that are related to lowering operators $\mathit{L}_{2\mp}=\ket{1-}\bra{2+}$ in $\mathit{P}_{\pm}$ and $\mathit{L}_{2\pm}=\ket{1+}\bra{2-}$ in $\mathit{P}_{\mp}$ respectively, behave different as a function of detuning $(\Delta)$. Note that both coefficients are the same at $\Delta=0$, but split up when $\Delta$ increases. This means that decay from $\ket{2+}$ to $\ket{1-}$ will be bigger than $\ket{2-}$ to $\ket{1+}$, as the system departs from $\Delta=0$. It is straight forward to check that $k_{n\pm}$ and $k_{n\mp}$ decrease when increasing the number of excitations $n$, see Eq.~(\ref{a_coefficient}). Consequently, the biggest contribution coming from $\mathit{P}_{\pm}$ and $\mathit{P}_{\mp}$ is for $n=2$. Therefore, this figure also implies that at large detuning $\Delta \gg g$, the interchanging operators $\mathit{P}_{\pm}$ and $\mathit{P}_{\mp}$ can be neglected. We shall illustrate how to further simplify the Lindblad operator in this regime. \par

As a consequence of the elimination of interchanging operators for $\Delta \gg g$, the annihilation operator can be now written as $a\approx \mathit{P}_{-}+\mathit{P}_{+}$. Let us now focus on the products $\mathit{P}_+^{\dagger}\mathit{P}_-$ and $\mathit{P}_-^{\dagger}\mathit{P}_+$ that appears in the anti-commutator term $\{a^{\dagger}a,\rho\}$ of the master equation~\eqref{ME}. Note that $\{\mathit{P}_+^{\dagger}\mathit{P}_-, \rho\}$ and $\{\mathit{P}_-^{\dagger}\mathit{P}_+, \rho\}$ vanish outside the subspace $n=1$. In addition, in the $n=1$ subspace they oscillate as a function of $g$ (see Eq.~(\ref{pmaspmen2}) for further details), and they can be neglected through the aforementioned RWA. For the operators of the form $\mathit{P}_{+}\rho\mathit{P}_-^\dagger$ the same approach of the RWA holds. Moreover, if the initial state $\ket{\psi(0)}$ is $\ket{2-}$, in the absence of interbranch exchange opetarors like the one coming from the hopping Hamiltonian, means that $\mathit{P}_{+}\rho\mathit{P}_-^\dagger$ and $\mathit{P}_{-}\rho\mathit{P}_+^\dagger$ operators are always zero. 
 
Branch conserving terms $\mathit{P}_+^{\dagger}\mathit{P}_+$ and $\mathit{P}_-^{\dagger}\mathit{P}_-$ yield no exponential time-dependence, regardless the manifold $n$. Therefore, for $\Delta\gg g$ and considering long-lived atoms, the general Lindbladian operator in Eq.~(\ref{ME}) can be written in the polariton basis where LP and UP losses are decoupled, such that $\mathcal{L}_{c}[\rho]=\mathcal{L}_+[\rho]+\mathcal{L}_-[\rho]$ , where
\begin{eqnarray}\label{lindblad}
\mathcal{L}_{x}[\rho] &=&\frac{\gamma}{2}(2\mathit{P}_{x}\rho\mathit{P}_{x}^\dagger -\lbrace \mathit{P}_{x}^\dagger \mathit{P}_{x},\rho \rbrace).
\end{eqnarray}

Then, the detuning is responsible for an asymmetry in the absorption spectrum, and it can be increased to suppress polaritons interchange. In the next subsection we explore the effects of a time-dependent detuning. 

\subsection{Time-Dependent Detuning}

In this subsection we focus on a single JC system, where detuning can be externally controlled, e.g. via Stark shift. To begin with, let us consider the Hamiltonian in the interaction picture, $V_I=g(a^{\dag}\sigma e^{-i\Delta t}+\sigma^{\dag}ae^{i\Delta t})$, with $\Delta=\omega^a-\omega^c$. In the subspace expanded by the states $\{\ket{n,g},\ket{n-1,e}\}$, we introduce the operators, $S_+ =a^{\dag}\sigma$ and $S_-=\sigma^{\dag}a$. Then, the Hamiltonian can be written as
 \begin{equation}
V_I=2g(S_x\cos(\Delta t)+S_y\sin(\Delta t)), 
 \end{equation}
where we use the relations $S_{\pm}=S_x \pm iS_y$. Notice that we can eliminate the $S_x$ contribution by selecting $\Delta t=\pi(2m+1)/2$, with $m \in \mathbb{Z}$, that leads us to $V_{I(m)}=2g(-1)^mS_y$. This interaction is responsible for coherent interchange of polariton branches, such as $V_{I(m)}\ket{n-}=g(-1)^mi\sqrt{n}\ket{n+}$. Note that under the constraint for $\Delta t$, $V_{I(m)}$ is a time-independent Hamiltonian that induces oscillations from $\ket{n-}$ to $\ket{n+}$ and viceversa, which follows from ($m=0$)
\begin{equation}
e^{-iV_It}\ket{n-}=\cos(gt\sqrt{n})\ket{n-}-\sin(gt\sqrt{n})\ket{n+}. 
\end{equation}
Therefore, a time-dependent detuning can be also used to coherently control polaritons interchange while remaining in the same manifold.

To summarize, we have focused on four different mechanisms that allow polaritons interchange. Firstly, we consider cavity hopping. Here, for a hopping strength fulfilling $J\leq 0.1\,g$, the system remains in the initial branch. Secondly, we analyze both cavity and atomic driving. We found that when these couplings are small, that is, $\alpha=\Omega=0.1\,g$, there is no interchange. Nevertheless, in the strong coupling regime $\Omega=50\,g$ ($\alpha=0$), and large cavity detuning $\Delta_c=500\,g$, Rabi oscillations are observed in the $n=1$ manifold ($\ket{1-}\leftrightarrow \ket{1+}$). Thirdly, by considering downwards transitions, we found that for large detuning $\Delta\geq 10\,g$, the Lindbladian operator decouples the two branches, and no interchange is observed. Moreover, the approximation is better when starting from the lower branch (LP) due to the asymmetry in the decay process, which originates from $k_{\pm}$ and $k_{\mp}$ coefficients. Finally, we considered an externally controlled detuning (time-dependent) that originates oscillations between LP and UP. The latter, as we shall see in the next section, directly affects the dynamics of Mott Insulator- and Superfluid-like states. This is because it imposes a constraint on the variation of the detuning that, up to our best knowledge, has not been explored in this context. 

For completeness, we quantify the coherence between states $\ket{1-}$ and $\ket{1+}$ generated by each mechanism. The coherence is commonly defined as the $l_1$-norm $C(t)=\sum_{i\neq j}|\rho_{ij}(t)|$~\cite{Baumgratz}, where $\rho_{ij}(t)=\langle i|\rho(t) |j\rangle$ are the matrix elements of the system density operator. However, since we are not interested in the overall coherence but in the particular coherence between states $\vert 1-\rangle$ and $\vert 1+\rangle$ as a measure of the degree of coherent control corresponding to each mechanism, we define $C(t)=|\rho_{1+,1-}(t)|+|\rho_{1-,1+}(t)|$. For the hopping mechanism we trace out over one of the cavities. In Table~\ref{Table1} we display the four mechanisms for controlling polaritons, including the probability for interchange of polariton branches ($P_{1+,1-}=\mbox{Tr}[\rho|1+,1-\rangle\langle1+,1-|]$ and $P_{1+}=\mbox{Tr}[\rho |1+\rangle\langle1-|]$) and the coherence originated in the process.

\begin{table}
\caption{\label{Table1}The Table shows the four mechanisms for controlling polaritons described in the main text, the corresponding control parameters, the minimum number of cavities required, the initial conditions used in the simulations, the maximum coherence and the probability for having interchange of polariton branches.}
\begin{ruledtabular}
\begin{tabular}{cccc}
 \mbox{Mechanism} & \mbox{Control Par.}  & \mbox{Coherence} & \mbox{Interchange} \\
  \# \mbox{Cavities}  &  $|\Psi(0)\rangle$ & $C(t)$  & \\
\hline
\mbox{hopping}& $J \;  (\sim g)$   & \mbox{coherent} & $P_{1+,1-} \approx 0.2$ \\
$2$ & $|1-,1-\rangle$   & $C \approx 0.4$ &  \\
\mbox{driving}& $ \Omega \; (\sim 50 g)$   & \mbox{coherent} & $P_{1+} \approx 1.0$ \\
$1$ & $|1-\rangle$   & $C \approx 1.0$ &  \\
\mbox{relaxation}& $ \gamma \; (\sim g)$   & \mbox{incoherent} & $P_{1+} \approx 0$ \\
$1$ & $|2-\rangle$   & $C \approx 0.1$ &  \\
\mbox{modulation}& $ \Delta \; (=\pi/(2t))$   & \mbox{coherent} & $P_{1+} \approx 1.0$ \\
$1$ & $|1-\rangle$   & $C \approx 1.0$ &  \\
\end{tabular}
\end{ruledtabular}
\end{table}

In what follows, we consider a parameter space where cavity hopping and longitudinal relaxation do not interchange polaritons.

\section{Detuning-controlled non-equilibrium dynamics}\label{QPT}

\begin{figure*}
	\centering
	\includegraphics[width=15 cm]{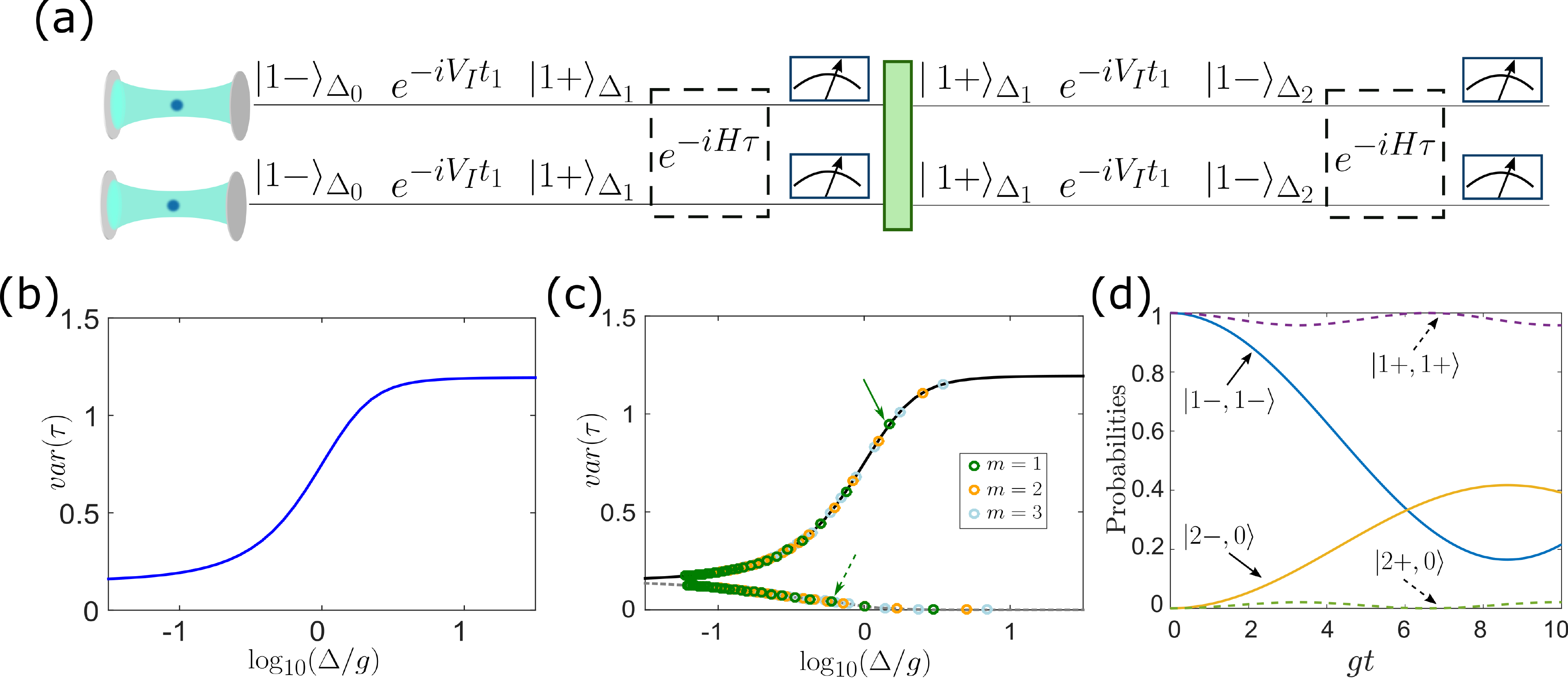}
	\caption{(a) Control sequence to measure the order parameter $var(\tau)$, while dynamically inducing polaritons interchange in the transition between Mott Insulator- and Superfluid-like states. The order parameter as a function of detuning,(b) without considering the time-dependency in detuning,(c) considering the time dependency. For the latter, $var(\tau)$ undergoes oscillations between LP (black-solid) and UP (grey-dashed). Three modes $m=1,2,3$ are considered for the time-dependent detuning $\Delta(t) = \pi (2m+1)/(2t)$.(d) Dynamics of relevant states for the points highlighted in Fig.~\ref{figure4}(c) corresponding to $m=1$. Other parameters are $\omega^c=10^{4}g$, $J=10^{-1}g$ and $\gamma=\kappa=0$. } 
\label{figure4}
\end{figure*}

In this section, we study the non-equilibrium dynamics of Mott Insulator- and Superfluid-like states considering the two-sites Jaynes-Cummings lattice as described in Fig.~\ref{figure1}(a) with atomic modulation given by $\Delta(t)$. The former, features polaritons placed in fixed lattice sites due to the low probability of hopping between neighboring sites. To explain this, suppose one cavity prepared in a state with one excitation of energy $E_{1-}$. Since the lowest energy for two excitations is $E_{2-}$, moving one additional excitation to the cavity requires an extra energy of, $E_{2-}-2E_{1-}=2\sqrt{g^2+\Delta^2/4} - \sqrt{2g^2+\Delta^2/4} - \Delta/2$, which plays the role of an effective one-site repulsion. Then, for $\Delta=0$ and $J/g\ll 1$, the atom-field interaction ($g$) on one site, shifts the frequency of the field causing a photon blockade effect \cite{Imamoglu,Birnbaum}. This repulsion can be tuned via detuning $\Delta$. The latter, features unbalanced distributions of polaritons across the lattice, as the above energy gap $E_{2-}-2E_{1-}$ tends to zero when increasing $\Delta$. Needless to say that our study shares a common ground with studies of Superfluid (SF) to Mott Insulator (MI) quantum phase transition (QPT)~\cite{Angelakis, Irish, Koch, Coto_2015,Hartmann06,Greentree,Greiner,Toyoda}. Therefore, some of our results could be extended to the QPT context. In order to hold this resemblance, we define a dynamical order parameter~\cite{Figueroa},%In this section, we introduce the quantum phase transition from the Mott Insulator (MI) to the superfluid (SF) state experienced by the JCH model \cite{Angelakis, Irish, Koch, Coto_2015,Hartmann06,Greentree,Greiner,Toyoda}. In particular, we consider the two-sites Jaynes-Cummings lattice as described in Fig.~\ref{figure1}(a) with atomic modulation given by $\Delta(t)$ and without atomic driving ($\omega_l$). In the MI phase, polaritons are placed in fixed lattice sites with a low probability of hopping between neighboring sites. To explain this, suppose one cavity prepared in a state with one excitation of energy $E_{1-}$. Since the lowest energy for two excitations is $E_{2-}$, moving one additional excitation to the cavity requires an extra energy of, $E_{2-}-2E_{1-}=2\sqrt{g^2+\Delta^2/4} - \sqrt{2g^2+\Delta^2/4} - \Delta/2$, which plays the role of an effective one-site repulsion. Then, for $\Delta=0$ and $J/g\ll 1$, the atom-field interaction ($g$) on one site, shifts the frequency of the field causing a photon blockade effect \cite{Imamoglu,Birnbaum}. This repulsion can be tuned via detuning $\Delta$. As the latter increases, the above energy gap $E_{2-}-2E_{1-}$ tends to zero, leading to the SF phase. The dynamical order parameter of the quantum phase transition (QPT) is \cite{Figueroa}
\begin{equation}\label{order_parameter_var}
var(\tau)=\sum_{i=1}^N 1/\tau\int_0^\tau (\mbox{Tr}[\hat{N}_i^2\rho(t)]-\mbox{Tr}[\hat{N}_i\rho(t)]^2)\, dt,
\end{equation}
with $\tau=1/J$ being the characteristic time scale for excitations exchange between resonators. $\hat{N}_i=a^\dagger_i a_i + \sigma^\dagger_i\sigma_i$ accounts for the number of excitations in the $i$th cavity. For MI-like states the number of polaritons per cavity is fixed, and this leads to a vanishing $var(\tau)$. In contrast, for SF-like states the number of local excitations fluctuates, leading to a non-vanishing variance. An important result of our work is that as we vary the detuning to change the gap (decrease or increase the on-site repulsion), this may induce interchange of polariton branches inside the cavity. For illustration, we start from an initial state with integer filling factor of one excitation per site, that is, $\ket{\psi(0)}=\ket{1-}\otimes\ket{1-}$. We set an initial large detuning $\Delta\approx 60 g$, and then we decrease it holding $\Delta t=\pi(2m+1)/2$. The sequence is illustrated in Fig.~\ref{figure4}(a). The time interval between two subsequent values of $\Delta$ is set to $gt_1=\pi/2$, which is shorter than the dynamics induced by the hopping Hamiltonian for $J=10^{-1}g$. In this regime, $t_1\ll 1/J$, we can consider an independent time evolution of each cavity governed by $V_I$ (without hopping). The Stark shift induces oscillations from $\ket{1-}$ to $\ket{1+}$ between two subsequent points ($\Delta_i$ and $\Delta_{i+1}$). After this, the whole system (including hopping) evolves for a time $\tau=1/J$ and right after we calculate the order parameter $var(\tau)$. In Fig.~\ref{figure4}, we show $var(\tau)$ as a function of ${\rm log}_{10}(\Delta/g)$ from two different approaches, (b) usual variation of detuning (time-independent) \cite{Angelakis,Irish,Koch}, (c) time-dependent detuning for $m=1,2,3$. The former, always exhibits a transition between MI- and SF-like states as the initial LP remains in the same brach. The latter, by properly choosing $\Delta t$ and $gt$, may remain in the lower branch (LP) and then experiences the same transition, or the system oscillates between the two branches, i.e. between LP and UP states, controlling the transition. Solid and dashed lines correspond to LP (black-solid) and UP (gray-dashed), respectively, for an evolution without considering the time-dependency of the detuning, i.e. there is no interchange of polariton branches. For further illustration, we show in Fig.~\ref{figure4}(d) the dynamics of the relevant states for the LP and UP highlighted in Fig.~\ref{figure4}(c). In the limit $\Delta\gtrsim g$, the LP occupies the Superfluid-like state $\ket{2-,0}$ while the UP remains in the Mott Insulator-like state $\ket{1+,1+}$. 

\begin{figure}[t]
\centering
\includegraphics[scale=0.118]{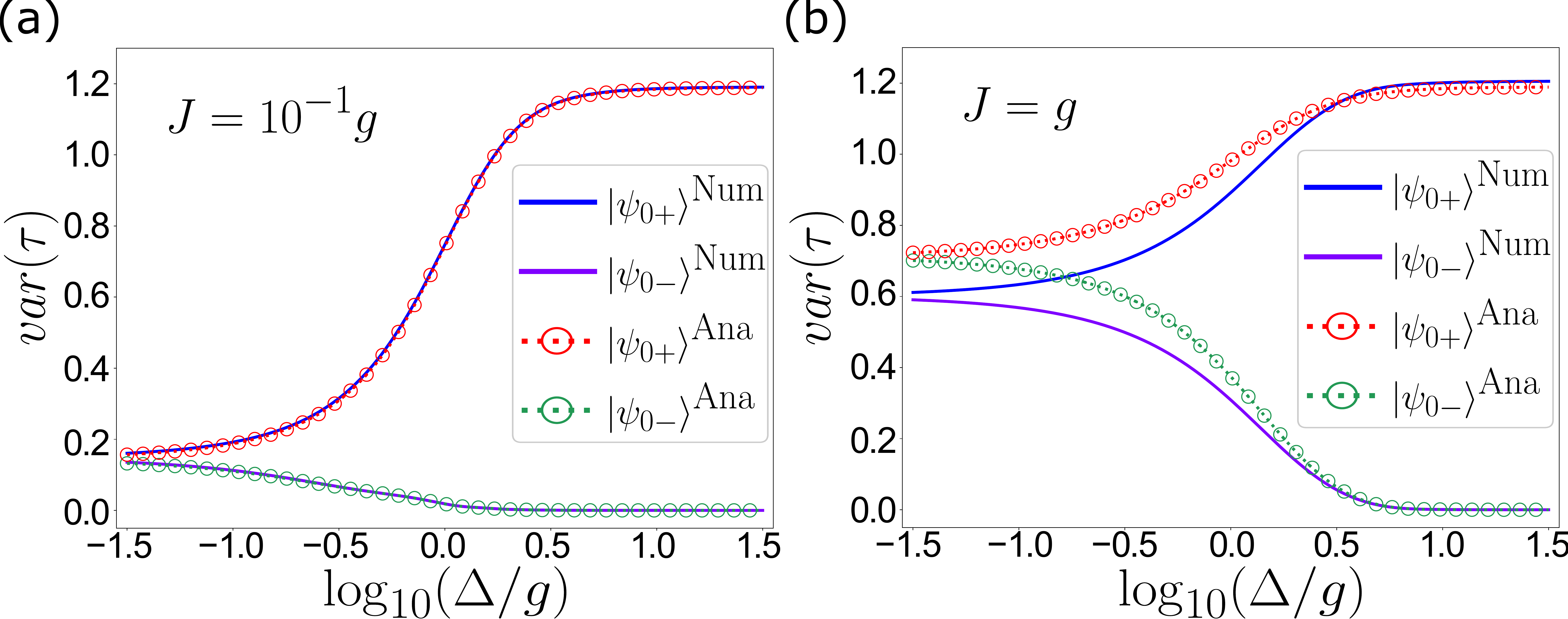}
\caption{(Color online) Approximated analytical expression for $var(\tau)$ agrees with the numerical calculations for the two initial conditions $\ket{\psi_{0+}}=\ket{1+,1+}$ and $\ket{\psi_{0-}}=\ket{1-,1-}$, in the regime $J\leq 10^{-1}g$. Other parameters are $\omega^c=10^{4}g$ and $\gamma=\kappa=0$.}
\label{figure6}
\end{figure}

For completeness, we seek for an analytical expression for $var(\tau)$. Following the approach in Ref.~\cite{Pena} and considering the quantum dynamics within the two-excitations manifold, given by $\{\ket{\psi_{0\pm}}=\ket{1\pm,1\pm}, \ket{\psi_1}\!\!=(\ket{2\pm,0}+\ket{0,2\pm})/\sqrt{2}\}$, the effective Hamiltonian reads,
\begin{equation}
H_{\rm eff}=
\left(
\begin{array}{cc}
a& b  \\
b & c  \\ 
\end{array}
\right).
\label{Effective1}
\end{equation}
For the initial condition $\ket{\psi_{0+}}=\ket{1+,1+}$, we set $a=2E_{1+}$, $b=-\sqrt{2}Jc_{1+}k_{2\pm}$, $c=2E_{2+}$, while for $\ket{\psi_{0-}}=\ket{1-,1-}$ we set $a=2E_{1-}$, $b=-\sqrt{2}Jc_{2-}c_{1-}$, $c=2E_{2-}$ where $c_{2-}c_{1-}=\cos(\theta_1)(\sqrt{2}\cos(\theta_1)\cos(\theta_2)+\sin(\theta_1)\sin(\theta_2))$, $k_{2\pm}c_{1+}=\sin(\theta_1)(\sqrt{2}\sin(\theta_1)\sin(\theta_2)+\cos(\theta_1)\cos(\theta_2))$. The full dynamics can be analytically solved  by diagonalizing the above  $2\times 2$ matrix. For instance, the time-averaged variance reads 
\begin{align}
var(\tau)=\frac{4b^2}{\Omega_0^2}\Bigg[1-\frac{J}{\Omega_0}\sin\bigg(\frac{\Omega_0}{J}\bigg)\Bigg],
\label{varianceI}
\end{align}
where we define $\Omega_0=\sqrt{4b^2+(a-c)^2}$. Worthwhile to notice that the key for our analytical expression relies on the separability of the polaritons. As shown in Fig.~\ref{figure4}(d), when the system is initialized in one branch, say the LP, the UP never shows up, and vice versa. Hence, whenever the polaritons start mixing our expression breaks down. We numerically found that this occur in the regime $J\approx g$, as shown in Fig.~\ref{figure6}.

\section{Conclusions}\label{Conclusions}

We explored several mechanisms for the interchange of polariton branches, implemented in a cavity QED lattice. Our results provide new insights about the regime where the hopping dynamics stemming from the Jaynes-Cumming-Hubbard model and losses originated from imperfect mirrors induce polariton interchange. Furthermore, we propose two mechanisms to coherently control Rabi oscillations between the lower and upper polariton branches in the one-excitation manifold. The first mechanism is based on atomic (two-level system) driving that induces oscillations in a second order process. The second one is based on atomic modulation that comes from a time-dependent detuning. We found that constraining the detuning to follow a specific evolution leads to heralded control of the transition between Mott Insulator- and Superfluid-like states. This result departs from the well-known observation of the order parameter for a time-independent detuning in the context of quantum phase transition. Moreover, when scaling the lattice's size, the control via time-dependent detuning can be used to modify transport's properties of the lattice. Finally, we study the role of detuning and hopping in the absorption spectrum of a cavity.

%\section{Funding}
%Fondo Nacional de Desarrollo Científico y Tecnológico (11180143,1160639,1190727).

\section{Acknowledgements}
We are grateful to M. Orszag and V. Eremeev for fruitful discussions. D.T and A.N. acknowledge financial support from Universidad Mayor through the Doctoral and Postdoctoral fellowships, respectively. R. Pe\~{n}a acknowledges the support from Vicerrector\'ia de Postgrado USACH. G.R. acknowledges the support from FONDECYT under grant No.1190727. F. T. acknowledges financial support from grants FA9550-16-1-0122, FA9550-18-1-0438, FONDECYT 
1160639, and CEDENNA through the Financiamiento Basal para Centros Cient\'ificos y Tecnológicos de 
Excelencia-FB0807. R.C. acknowledges financial support from FONDECYT Iniciación No. 11180143.

\section{Disclosures}
The authors declare no conflicts of interest.

\appendix

\section{Coefficients in the polariton basis}\label{Coefficients}

The coefficients in Eqs.~\eqref{adagger} and \eqref{sigma_dagger} are, for $n=1$, $c_{1+}=\sin(\theta_1)$, $c_{1-}=\cos(\theta_1)$, $c_{1+}^a=\cos(\theta_1)$, $c_{1-}^a=-\sin(\theta_1)$, and for $n \geq 2$
\begin{eqnarray}\label{a_coefficient}
c_{n+}&=&\sqrt{n}\sin(\theta_n)\sin(\theta_{n-1})+\sqrt{n-1}\cos(\theta_n)\cos(\theta_{n-1}) \hspace*{0.2cm}, \nonumber  \\
c_{n-}&=&\sqrt{n}\cos(\theta_n)\cos(\theta_{n-1})+\sqrt{n-1}\sin(\theta_n)\sin(\theta_{n-1}) \hspace*{0.2cm} ,   \nonumber \\
k_{n\pm}&=&\sqrt{n}\sin(\theta_n)\cos(\theta_{n-1})-\sqrt{n-1}\cos(\theta_n)\sin(\theta_{n-1}) \hspace*{0.2cm}, \nonumber \\
k_{n\mp}&=&\sqrt{n}\cos(\theta_n)\sin(\theta_{n-1})-\sqrt{n-1}\sin(\theta_n)\cos(\theta_{n-1}) \hspace*{0.2cm},\nonumber\\
\end{eqnarray}
and 
\begin{eqnarray}\label{sigma_coefficient}
c_{n+}^a &=&\cos(\theta_{n})\sin(\theta_{n-1}), \nonumber \\%\hspace*{0.2cm} n\geq 2, \nonumber \\
c_{n-}^a &=&-\sin(\theta_{n})\cos(\theta_{n-1}), \nonumber \\ %\hspace*{0.2cm} n\geq 2,  \nonumber \\
k_{n\pm}^a &=&\cos(\theta_{n})\cos(\theta_{n-1}), \nonumber \\% \hspace*{0.2cm} n\geq 2, \nonumber \\
k_{n\mp}^a &=&-\sin(\theta_{n})\sin(\theta_{n-1}),%\hspace*{0.2cm} n\geq 2. 
\end{eqnarray}
where  $\theta_{n}=\frac{1}{2}\arctan(\frac{g\sqrt{n}}{\Delta/2})$, $\Delta=\omega^a-\omega^c$, and $n$ corresponds to the number of photons inside each cavity.

\section{Transformation of the Hopping Hamiltonian}\label{Appendix1}

Let's consider the unitary operation $U=e^{iH_{\rm JC} t}$ and expand it as,
\begin{equation}\label{U}
U = \prod_{j=1}^{N_{f}}\prod_{j^\prime =1}^{N_{f}} e^{it\sum_{n=1}^{N_{f}} E_{n+}^j\vert n+\rangle_j\langle n+\vert} e^{it\sum_{{n}^\prime =1}^{N_{f}} E_{{n}^\prime -}^{j^\prime}\vert {n}^\prime -\rangle_{j^\prime}\langle {n}^{\prime}-\vert}.
\end{equation}

From the above equation one can see that it is possible to separate the unitary transformation $U$ into two unitary transformations, one for each branch, $U=U_+U_-$. Let's focus first on the simple case where projectors leave the system in the same manifold, e.g. $\vert n_\pm\rangle\langle n_\pm\vert$ and $\vert n_\pm\rangle\langle n_\mp\vert$. The first projector do not transform under $U$, due to orthogonal relations, $\langle n_+\vert n_-\rangle =0$ and $\langle n_\pm\vert n_\pm^\prime\rangle =\delta_{n,n^\prime}$. For the second one, we use the relation
\begin{equation}\label{theorem}
e^{\beta A}Be^{-\beta A}=B+\beta[A,B]+\frac{\beta^2}{2!}[A,[A,B]]+\dots,
\end{equation}
which will be useful for all the calculations. Since operators for different cavities commute, we will omit the index $j$ and $j^\prime$. Then,
\begin{eqnarray}
U\vert n+\rangle\langle n-\vert U^{\dagger}&=& U_-U_+\vert n+\rangle\langle n-\vert U_+^{\dagger}U_-^{\dagger}\nonumber\\&=&e^{it(E_{n+}-E_{n-})}\vert n+\rangle\langle n-\vert ,
\end{eqnarray} 
and the exponent $E_{n+}-E_{n-}=\sqrt{\Delta^2 +4g^2n}$.

We now focus on the hopping Hamiltonian in Eq.~(\ref{Hhp_pol}). After doing the products, we transform each operator separately, e.g. $UP_{+j}^{\dagger}P_{+(j+1)}U^{\dagger}=\tilde{P}_{+j}^{\dagger}\tilde{P}_{+(j+1)}$. For instance,
\begin{equation}\label{L_mas_dagger}
U\mathit{L}_{n+}^{\dagger}U^\dagger=U\vert n+\rangle\langle(n-1)+\vert U^\dagger = \mathit{L}_{n+}^{\dagger} e^{it(E_{n+}-E_{(n-1)+})} ,
\end{equation}
where we have used Eq.~(\ref{theorem}). For these kind of projectors we must perform only one transformation, say $U_+$, since $U_-$ commutes with the projector. The exponent $E_{n+}-E_{(n-1)+}=\omega^c +\frac{1}{2}(R_n - R_{n-1})$, with $R_{n}=\sqrt{\Delta^2+4g^2n}$. Therefore, the hopping interaction is
\begin{eqnarray}\label{pmas1}
\tilde{P}_{+j}^{\dagger}\tilde{P}_{+(j+1)}&=&\sum_{n=1}^{N_{f}} c_{n+}\mathit{L}_{n+}^{j\dagger} e^{it(R_{n}-R_{(n-1)})} \nonumber\\&\times&\sum_{n^\prime =1}^{N_{f}} c_{n^\prime +}\mathit{L}_{n^\prime +}^{(j+1)} e^{-it(R_{n^\prime}-R_{(n^\prime -1)})}.
\end{eqnarray}

For $\tilde{P}_{-j}^{\dagger}\tilde{P}_{-(j+1)}$, we simply replace $n+ (n^\prime+)\rightarrow n-(n^\prime-)$ in Eq.(\ref{pmas1}). It is worth noticing that in the manifold $n=1$ both $\tilde{P}_{+j}^{\dagger}\tilde{P}_{+(j+1)}$ and $\tilde{P}_{-j}^{\dagger}\tilde{P}_{-(j+1)}$ cancel the exponential dependence with $R_n$, henceforth these operators cannot be eliminated under a Rotating Wave Approximation (RWA).

For the product $P_{+j}^{\dagger}P_{-(j+1)}$ we get,
\begin{eqnarray}\label{pmaspmen2}
\tilde{P}_{+j}^{\dagger}\tilde{P}_{-(j+1)}&=&\sum_{n=1}^{N_{f}} c_{n+}\mathit{L}_{n+}^{j\dagger} e^{it(R_{n}-R_{(n-1)})}\nonumber\\&\times&\sum_{n^\prime =1}^{N_{f}} c_{n^\prime -}\mathit{L}_{n^\prime -}^{(j+1)} e^{it(R_{n^\prime }-R_{(n^\prime -1)})}.
\end{eqnarray}

Note that $\tilde{P}_{+j}^{\dagger}\tilde{P}_{-(j+1)}$ always oscillates with frequency proportional to $R_n$, and thus it can be eliminated under the RWA. For operators of the form $P_{\pm}$, let's calculate first $\mathit{L}_{n\pm}^\dagger$, 
\begin{equation}
U \mathit{L}_{n\pm}^\dagger U^{\dagger} = U\vert n+\rangle\langle(n-1)-\vert U^\dagger = \mathit{L}_{n\pm}^\dagger e^{it(E_{n+}-E_{(n-1)-})},
\end{equation}
where the exponent $E_{n+}-E_{(n-1)-}=\omega^c +\frac{1}{2}(R_{n} + R_{(n-1)})$. Then,
\begin{eqnarray}\label{pmas_pmame}
\tilde{P}_{+j}^{\dagger}\tilde{P}_{\pm(j+1)}&=&\sum_{n=1}^{N_{f}} c_{n+}\mathit{L}_{n+}^{j\dagger} e^{it(R_{n}-R_{(n-1)})}\nonumber\\&\times&\sum_{n^\prime =1}^{N_{f}} k_{n^\prime \pm}\mathit{L}_{n^\prime \mp}^{(j+1)} e^{it(R_{(n^\prime -1 )}+R_{n^\prime})},
\end{eqnarray}
and
\begin{eqnarray}\label{pmas_pmema}
\tilde{P}_{+j}^{\dagger}\tilde{P}_{\mp(j+1)}&=&\sum_{n=1}^{N_{f}} c_{n+}\mathit{L}_{n+}^{j\dagger} e^{it(R_{n}-R_{(n-1)})}\nonumber\\&\times&\sum_{n^\prime =1}^{N_{f}} k_{n^\prime \mp}\mathit{L}_{n^\prime \pm}^{(j+1)} e^{-it(R_{n^\prime}+R_{(n^\prime-1)})}.
\end{eqnarray}

The above operators (Eqs.~\eqref{pmas_pmame}-\eqref{pmas_pmema}) vanish in the manifold $n=1$, since $k_{1\pm}=k_{1\mp}=0$. Moreover, for $n\geq2$ these operators oscillate in time and they can be eliminated under the RWA. Finally, operators like $\tilde{P}_{\mp j}^{\dagger}\tilde{P}_{\mp(j+1)}$, have a small contribution because the quadratic dependence with $k_{n}$.

\section{Perturbation theory}\label{Appendix3}

At first order there is no correction for any of the eigenvalues, which can be rapidly notice from the absence of diagonal elements in the perturbative terms (those proportional to $\beta$ and $\xi$). Then

\begin{equation}
E_{k}^{(1)}=\bra{k^{(0)}}\tilde{H}_i\ket{k^{(0)}}=0,
\end{equation}

with $k=\lbrace G,1-,1+,2-,2+ \rbrace$ the unperturbed eigenstate of $\tilde{H}_0$. The ground state $\ket{G}$, with zero eigenvalue ($E_G^{(0)}=0$) has been included as well. For the eigenstate the corrections at first order reads
\begin{flalign}
&\ket{G}^{(1)} = -a_+ \ket{1+}-a_- \ket{1_-},\\
&\ket{1-}^{(1)} = -a_-\ket{G}+ b_-\ket{2-}+ c_{-,\pm}\ket{2+}  ,\\
&\ket{1+}^{(1)} = -a_+\ket{G}+ b_+\ket{2+}+ c_{+,\mp}\ket{2-} ,\\
&\ket{2-}^{(1)} =  -b_- \ket{1-} - c_{+,\mp}\ket{1+} + d_-\ket{3-}+e_{-,\pm}\ket{3+},\\
&\ket{2+}^{(1)} = -b_+ \ket{1+} - c_{-,\pm}\ket{1-} + d_+\ket{3+}+e_{+,\mp}\ket{3-},
\end{flalign}
where
\begin{flalign}
a_{\eta} &= {\beta_{1\eta} \over E_{1\eta}^{(0)}}, \; b_{\eta} = {\beta_{2\eta} \over E_{1\eta}^{(0)}-E_{2-\eta}^{(0)}}, \;  d_{\eta} = {\beta_{3\eta} \over E_{3\eta}^{(0)}-E_{3\eta}^{(0)}},\\
c_{\eta,\mu} &= {\xi_{2\mu} \over E_{1\eta}^{(0)}-E_{2-\eta}^{(0)}}, \;
e_{\eta,\mu}= {\xi_{3\mu} \over E_{2\eta}^{(0)}-E_{3\eta}^{(0)}},
\end{flalign}
with $\eta = +,-$, $\mu = \pm, \mp$, and $-(\mp) = \pm$. The corrections to the eigenvalues at second order are explicitly given by
\begin{eqnarray}\label{2nd_ener_correct}
E_{G}^{(2)} &=& \sum_{k\neq G} \frac{\vert\bra{k^{(0)}}\tilde{H}_i\ket{G}\vert ^2}{E_G^{(0)} - E_{k}^{(0)}}=-\frac{\vert \beta_{1-}\vert^2}{E_{1-}^{(0)}} - \frac{\vert \beta_{1+}\vert^2}{E_{1+}^{(0)}}	,\nonumber\\ 
E_{1-}^{(2)} &=& \frac{\vert - \beta_{1-}\vert^2}{E_{1-}^{(0)}} + \frac{\vert \beta_{2-}\vert^2}{E_{1-}^{(0)} - E_{2-}^{(0)}} + \frac{\vert \xi_{2\pm}\vert^2}{E_{1-}^{(0)} - E_{2+}^{(0)}}	,\nonumber\\
E_{1+}^{(2)} &=& \frac{\vert - \beta_{1+}\vert^2}{E_{1+}^{(0)}} + \frac{\vert \beta_{2+}\vert^2}{E_{1+}^{(0)} - E_{2+}^{(0)}} + \frac{\vert \xi_{2\mp}\vert^2}{E_{1+}^{(0)} - E_{2-}^{(0)}}	,\nonumber\\
E_{2-}^{(2)} &=& \frac{\vert - \beta_{2-}\vert^2}{E_{2-}^{(0)}-E_{1-}^{(0)}} + \frac{\vert - \xi_{2\mp}\vert^2}{E_{2-}^{(0)} - E_{1+}^{(0)}} + \frac{\vert \beta_{3-}\vert^2}{E_{2-}^{(0)} - E_{3-}^{(0)}} \nonumber \\ &+& \frac{\vert \xi_{3\pm}\vert^2}{E_{2-}^{(0)} - E_{3+}^{(0)}}	,\nonumber\\
E_{2+}^{(2)} &=& \frac{\vert - \beta_{2+}\vert^2}{E_{2+}^{(0)}-E_{1+}^{(0)}} + \frac{\vert - \xi_{2\pm}\vert^2}{E_{2+}^{(0)} - E_{1-}^{(0)}} + \frac{\vert \beta_{3+}\vert^2}{E_{2+}^{(0)} - E_{3+}^{(0)}} \nonumber \\ &+& \frac{\vert \xi_{3\mp}\vert^2}{E_{2+}^{(0)} - E_{3-}^{(0)}}	.
\end{eqnarray}

Similarly, for the eigenstates we get 
\begin{eqnarray}
\ket{G}^{(2)} &=& \sum_{k,l\neq G}\frac{\bra{k^{(0)}}\tilde{H}_i\ket{l^{(0)}}\bra{l^{(0)}}\tilde{H}_i\ket{G}}{(E_G^{(0)} - E_{k}^{(0)})(E_G^{(0)} - E_{l}^{(0)})} \ket{k^{(0)}} \nonumber \\ 
&& =\left(\frac{\beta_{1-}\beta_{2-}}{E_{1-}^{(0)} E_{2-}^{(0)}} + \frac{\beta_{1+}\xi_{2\mp}}{E_{1+}^{(0)} E_{2-}^{(0)}}\right)\ket{2-}\nonumber\\
&& + \left(\frac{\beta_{1+}\beta_{2+}}{E_{1+}^{(0)} E_{2+}^{(0)}} + \frac{\beta_{1-}\xi_{2\pm}}{E_{1-}^{(0)} E_{2+}^{(0)}} \right)\ket{2+},
\end{eqnarray}
and for the other states we found the following compact form
\begin{eqnarray}
\ket{1-}^{(2)} &=& f_{-,\mp}|1+\rangle+g_{-,\mp}|3-\rangle+h_{-,\pm}|3+\rangle,\\
\ket{1+}^{(2)} &=& f_{+,\pm}|1+\rangle+g_{+,\pm}|3-\rangle+h_{-,\mp}|3+\rangle,\\
\ket{2-}^{(2)} &=& i_{-,\mp}|2+\rangle + j_{-,\mp}|G\rangle,\\
\ket{2+}^{(2)} &=& i_{+,\pm}|2-\rangle + j_{+,\pm}|G\rangle,
\end{eqnarray}
where the coefficients are defined as
\begin{eqnarray}
f_{\eta,\mu} &=& {\displaystyle{-\frac{\beta_{1\eta}\beta_{1-\eta}}{E_{1\eta}^{(0)}} - \frac{\beta_{2\eta}\xi_{2\mu}}{E_{1\eta}^{(0)}- E_{2\eta}^{(0)}} - \frac{\xi_{2-\mu}\beta_{2-\eta}}{E_{1\eta}^{(0)} - E_{2-\eta}^{(0)}}} \over E_{1\eta}^{(0)} - E_{1-\eta}^{(0)}}, \\
g_{\eta,\mu} &=& {\displaystyle{\frac{\beta_{2\eta}\beta_{3\mu}}{E_{1\eta}^{(0)}- E_{2\eta}^{(0)}} + \frac{\xi_{2 -\mu}\xi_{3\mu}}{E_{1\eta}^{(0)}- E_{2-\eta}^{(0)}}} \over E_{1\eta}^{(0)} - E_{3-\eta}^{(0)}}, \\
h_{\eta,\mu} &=& {\displaystyle{\frac{\beta_{2\eta}\xi_{3\mu}}{E_{1\eta}^{(0)}- E_{2\eta}^{(0)}} + \frac{\xi_{2 \mu}\beta_{3-\eta}}{E_{1\eta}^{(0)}- E_{2-\eta}^{(0)}}} \over E_{1\eta}^{(0)} - E_{3-\eta}^{(0)}}, \\
i_{\eta,\nu} &=& {1 \over E_{2\eta}^{(0)} - E_{2-\eta}^{(0)}}\left(-\frac{\beta_{2\eta}\xi_{2-\nu}}{E_{2\eta}^{(0)}-E_{1-\eta}^{(0)}} - \frac{\xi_{2\nu}\beta_{2-\eta}}{E_{2\eta}^{(0)}- E_{1-\eta}^{(0)}} \right. \nonumber \\
&&\left. - \frac{\beta_{3\eta}\xi_{3\nu}}{E_{2\eta}^{(0)} - E_{3-\eta}^{(0)}} -\frac{\xi_{3-\nu}\beta_{3-\eta}}{E_{2\eta}^{(0)} - E_{3-\eta}^{(0)}} \right), \\
j_{\eta,\mu} &=& {1 \over E_{2\eta}^{(0)}}\left(\frac{\beta_{2\eta}\beta_{1\eta}}{E_{2\eta}^{(0)}- E_{1\eta}^{(0)}} + \frac{\xi_{2\mu}\beta_{1-\eta}}{E_{2\eta}^{(0)}- E_{1-\eta}^{(0)}}\right).
\end{eqnarray}

\section{Absorption spectrum}\label{AbsorptionSpectrum}

Let us consider a single cavity QED governed by the Markovian master equation $\dot{\rho} = \mathcal{L}[\rho]$, where $\mathcal{L}$ and $\rho(t)$ are the Lindbladian and density matrix of the system, respectively. When photonic and atomic losses are considered, we have $\mathcal{L}[\rho] = -i[H_{\rm JC},\rho]+\mathcal{L}_a [\rho] + \mathcal{L}_{\sigma_-}[\rho]$, where $H_{\rm JC}$ is the Jaynes-Cummings Hamiltonian~\eqref{Hjc} for $N_c = 1$, and the two dissipation channels are described by
\begin{eqnarray}
\mathcal{L}_a [\rho] &=& \frac{\gamma}{2}(2a\rho a^\dagger -\lbrace a^\dagger a,\rho \rbrace), \\
\mathcal{L}_{\sigma_-} [\rho] &=& \frac{\kappa}{2}(2\sigma_{-}\rho \sigma_-^\dagger -\lbrace \sigma_-^\dagger  \sigma_-,\rho \rbrace),
\end{eqnarray}
where $\gamma$ and $\kappa$ are the photonic and atomic decay rates, respectively. If a pumping laser with frequency $\omega$ weakly drives the system, the absorption spectrum can be defined as the Fourier transform of the photonic two-point correlation function $G(\tau) = \langle \langle a(\tau)a^{\dagger}(0)\rangle \rangle_{\rm ss}$,
\begin{equation}
S(\omega) = 2\mbox{Re} \int_{0}^{\infty} G(\tau) e^{i\omega \tau}\; d\tau.
\end{equation}

The double expectation value means deviations with respect its stationary state, i.e. $G(\tau) = \langle a(\tau)a^{\dagger}(0) \rangle_{\rm ss} - \lim_{\tau \rightarrow \infty}\langle a(\tau)a^{\dagger}(0) \rangle_{\rm ss}$ with $\langle a(\tau)a^{\dagger}(0)\rangle \rangle_{\rm ss} = \mbox{Tr}(a(\tau)a^{\dagger}(0)\rho_{\rm ss})$~\cite{Neuman}. Here, $\rho_{\rm ss}$ is the steady state (ss) of the system which can be found by solving the condition
\begin{equation}
\mathcal{L} [\rho_{\rm ss}] = 0.
\end{equation}

To numerically find $\rho_{\rm ss}$ we solves the eigenvalue equations $\mathcal{L}[R_k] = \lambda_k R_k$ and $\mathcal{L}^{\dagger}[L_k] = \lambda_k L_k$, where $R_k$($L_k$) and $\lambda_k$ are the right(left) eigenmatrices and eigenvalues, respectively. As the general solution is given by $\rho(t) = \sum_k c_k e^{\lambda_k t}R_k$, where $c_k = \mbox{Tr}(\rho(0)L_k)$~\cite{Dominic2016,Katarzyna2016}, from the zero eigenvalue $\lambda_0 = 0$, we compute $\rho_{\rm ss} = c_0 R_0$. On the other hand, the expectation value $\langle a(\tau)a^{\dagger}(0) \rangle_{\rm ss}$ is calculated using the quantum regression theorem~\cite{Breuerbook}, as follow
\begin{equation}
\langle a(\tau)a^{\dagger}(0) \rangle_{\rm ss} = \mbox{Tr}\left[a(0)f(\tau) \right],
\end{equation}
where $f(\tau) = e^{\mathcal{L}\tau}(a^{\dagger}(0)\rho_{\rm ss})$ satisfy the master equation
\begin{equation}
\dot{f} = \mathcal{L}[f], \quad  \quad f(0) = a^{\dagger}(0)\rho_{\rm ss},
\end{equation}
with $a^{\dagger}(0)\rho_{\rm ss}$ is the initial condition of the function $f(\tau)$. To numerically compute $f(t)$ we use the standard general solution of the Lindblad master equation~\cite{Norambuena_2020}.

In the three-level manifold composed by the states $|1\rangle = |1+\rangle$, $|2\rangle = |1-\rangle$, and $|3\rangle = |0g\rangle\}$ photonic and atomic operators takes the form $a = \sin(\theta_1)|3\rangle \langle 1|+\cos(\theta_1)|3\rangle \langle 2|$ and $\sigma^{-} = \cos(\theta_1)|3\rangle \langle 1|-\sin(\theta_1)|3\rangle \langle 2|$, respectively. The two-point correlation function can be calculated using the three-level picture, resulting in $G(\tau) = \sin(\theta_1)[f_{13}(\tau)-f_{13}(\infty)]+\cos(\theta_1)[f_{23}(\tau)-f_{23}(\infty)]$, where $f_{ij} = \langle i|f|j\rangle$. From the master equation $\dot{f} = \mathcal{L}[f]$, where $\mathcal{L}$ is the Lindbladian given in~\eqref{ME} it follows that $f_{ij}(\tau) = f_{ij}(0)\mbox{exp}[(-i\omega_{ij}-\gamma_{ij})\tau]$, where $\omega_{ij} = (E_i-E_j)/\hbar$, $E_{i}$ are the polaritonic energies,$\gamma_{13} = (1/2)(\sin^2(\theta_1)\gamma +\cos^2(\theta_1)\kappa) = \gamma_{+}$ and $\gamma_{23} = (1/2)(\cos^2(\theta_1)\gamma +\sin^2(\theta_1)\kappa)= \gamma_{-}$ are the decay rates of polaritonic states $E_{1+}$ and $E_{-1}$, respectively. Using these results we reproduce the analytical expression given in Eq.~\eqref{AbsSpectrumAn}.

\end{document}